\documentclass{PoS}
\usepackage{caption}
\usepackage{graphicx}
\usepackage{citesort}

\usepackage{wrapfig}
\usepackage[utf8]{inputenc}

\makeatletter 
\renewcommand\speaker[1]{\if@speaker\global\@dblspeaktrue\fi
			\global\@speakertrue
			\global\setbox\@firstaubox
			\hbox{{\let\thanks\@gobble
				\let\footnote\@gobble\small
				\rm  The nCTEQ Collaboration}}%
			#1\thanks{Speaker.}\
			}%
\makeatother

\title{LHC data and its impact on nCTEQ15 PDFs}

\ShortTitle{LHC data and its impact on nCTEQ15 PDFs}

\author{%
The  nCTEQ Collaboration:\thanks{%
We acknowledge the hospitality of CERN, DESY, and Fermilab where a
portion of this work was performed.
This work was also partially supported by the U.S.\ Department of
Energy under Grant No.\ DE-SC0010129.
}\qquad
D.~B.~Clark\rlap,${}^1$ 
E.~Godat\rlap,${}^1$ 
T.~Je\v{z}o\rlap,${}^2$ 
C.~Keppel\rlap,${}^3  $
K.~Kova\v{r}\'{i}k\rlap,${}^4$ 
A.~Kusina\rlap,${}^{5,6}$ 
F.~Lyonnet\rlap,${}^1$ 
J.G.~Morfin\rlap,${}^7$
F.~I.~Olness\rlap,${}^1$\speaker{}
J.F.~Owens\rlap,${}^8$
I.~Schienbein\rlap,${}^5$ 
J.~Y.~Yu${}^1$ 
\\
${}^1$Southern Methodist University, Dallas, TX 75275, USA\\ 
${}^2$Physik-Institut, Universit\"at Z\"urich, Winterthurerstrasse 190, CH-8057 Z\"urich,
Switzerland\\
${}^3$Thomas Jefferson National Accelerator Facility, Newport News, VA, 23606, USA\\
${}^4$Institut f\"{u}r Theoretische Physik, Westf\"{a}lische Wilhelms-Universit\"{a}t
M\"{u}nster, \\ \qquad
Wilhelm-Klemm-Stra{\ss}e 9, D-48149 M\"{u}nster, Germany \\
${}^5$Laboratoire de Physique Subatomique et de Cosmologie, Universit\'{e}
Grenoble-Alpes, \\ \qquad
CNRS/IN2P3, 53 avenue des Martyrs, 38026 Grenoble,
France \\
${}6$Institute of Nuclear Physics, Polish Academy of Sciences, \\ \qquad
ul. Radzikowskiego 152, 31-342 Cracow, Poland\\
${}^7$Fermi National Accelerator Laboratory, Batavia, Illinois 60510, USA\\
${}^8$Department of Physics, Florida State University, Tallahassee, Florida 32306-4350, USA\\
}

\abstract{
The LHC heavy ion data for $W/Z$ production 
can provide new incisive information on the PDFs. 
This data is sensitive to the 
heavier quark flavors (strange  and charm) 
in a  high energy kinematic region; 
this can
facilitate the determination of PDFs 
in the  small $x$ region where previous data was limited. 
At present, the flavor separation of the proton PDFs is
dependent on DIS data from nuclear targets. 
Therefore, 
improved nuclear corrections can also yield
enhanced  flavor determination of both 
the proton and nuclear PDFs. 
}

\FullConference{XXV International Workshop on Deep-Inelastic Scattering and Related Subjects\\
		3-7 April 2017\\
		University of Birmingham, UK}

\begin{document}
%

\section{Introduction}
\nocite{Kovarik:2015cma}


\begin{wrapfigure}{R}{0.45\textwidth} 
\centering{} 
\vspace{-45pt}
\includegraphics[width=0.45\textwidth]{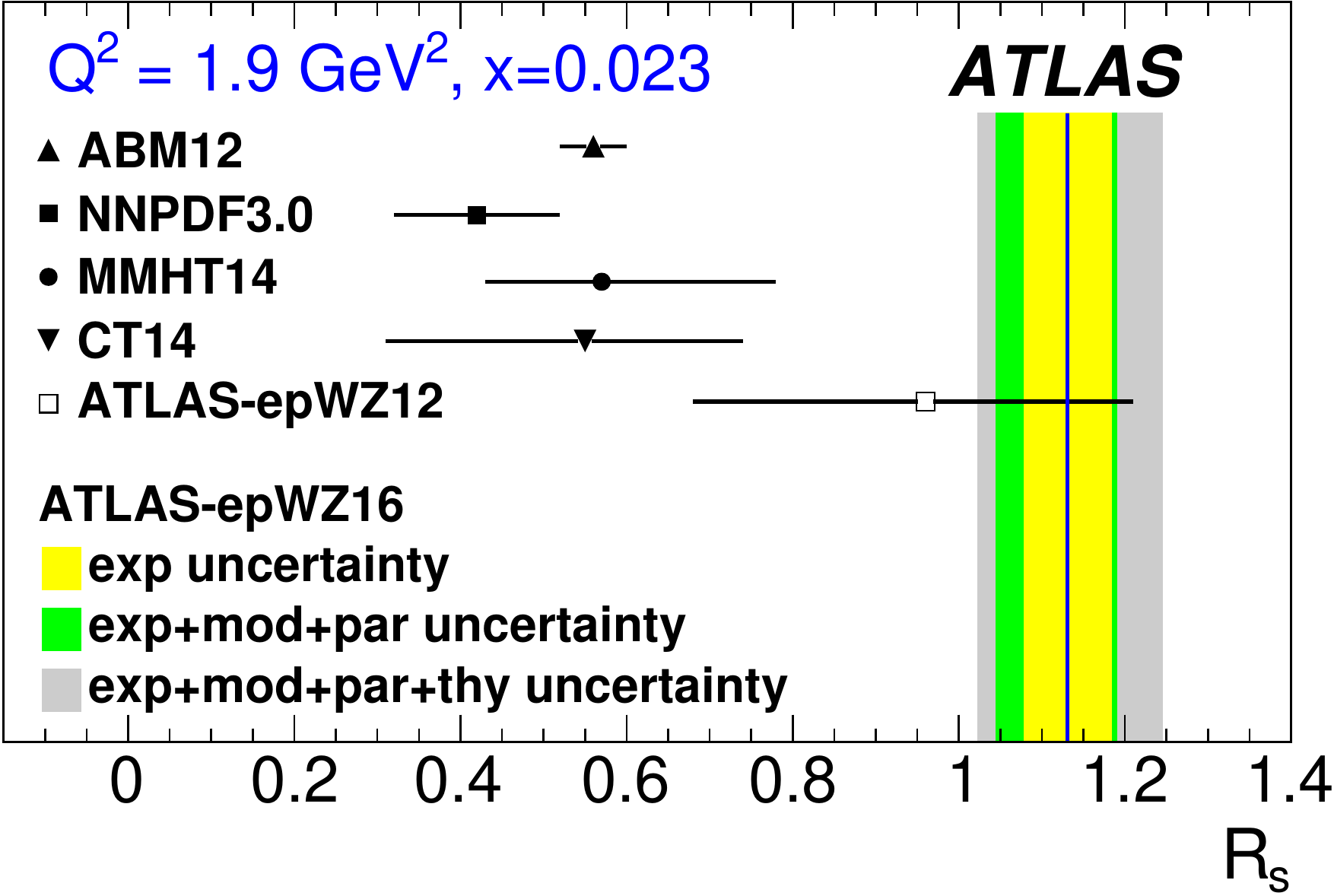}
\vspace{-20pt}
\caption{
The relative quark ratio $R_s={(s+\bar{s})/(\bar{u}+\bar{d})}$ 
as measured by the ATLAS collaboration from $W/Z$   production~\cite{Aaboud:2016btc}.
}
\label{fig:atlas}
\end{wrapfigure}

\begin{figure}[b] 
\centering{} 
\includegraphics[width=0.95\textwidth]{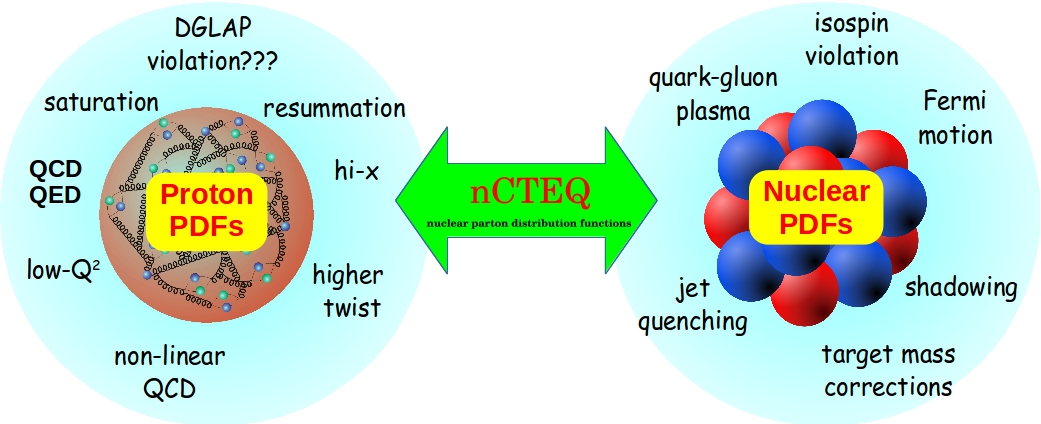}
\vspace{0pt}
\caption{Schematic representation of selected 
phenomenological issues 
that can impact the determination of proton and nuclear PDFs.\cite{Kovarik:2015cma}
}
\label{fig:scheme}
\end{figure} 

In May  2015, 
the LHC reached a record setting energy of 13~TeV for proton-proton collisions; 
it is  likely the LHC  will continue to hold the energy record for the foreseeable future 
until  it is surpassed by a FCC  facility. 
Therefore, our search for ``new physics'' phenomena turns to a careful analysis  of 
the accumulating LHC data to make incisive comparisons between experimental measurements and theoretical predictions.
The Parton Distribution Functions (PDFs) are the key ingredient that enables 
us to connect experiment with theory by applying  the QCD improved parton model to describe the  
distribution of quarks and gluons in the proton.

Despite  decades of studies, there is yet much to learn about the proton structure. 
A very interesting new result which was presented at this meeting was the ratio of the 
strange PDF to the up- and down-sea quark PDFs, as shown in Fig.~\ref{fig:atlas}.
As the strange quark  is heavier than the up and down quarks, 
the  common expectation was that the strange PDF would be suppressed 
relative to the other sea-quarks. 
Instead, this measurement  suggests the proton may be closer to 
an SU(3) flavor symmetric structure ($\bar{u}=\bar{d}=\bar{s}$)
rather than a suppressed strange PDF ($\bar{s}<\bar{u},\bar{d}$).

The above case is just one  example where improved determinations  of the PDFs 
can yield superior precision  for our theoretical predictions,
and thus an enhanced ability to discern ``new physics'' signatures from uncertain ``standard model''  processes. 

The goal of the nCTEQ collaboration is to make maximal use of the available data,  both proton and nuclei, to 
obtain the most precise determination of the PDFs. In this brief report, we  
summarize some of the recent advancements toward this goal.

\begin{figure}[t] 
\centering{} 
\null \vspace{-0.20in}
\includegraphics[width=0.90\textwidth]{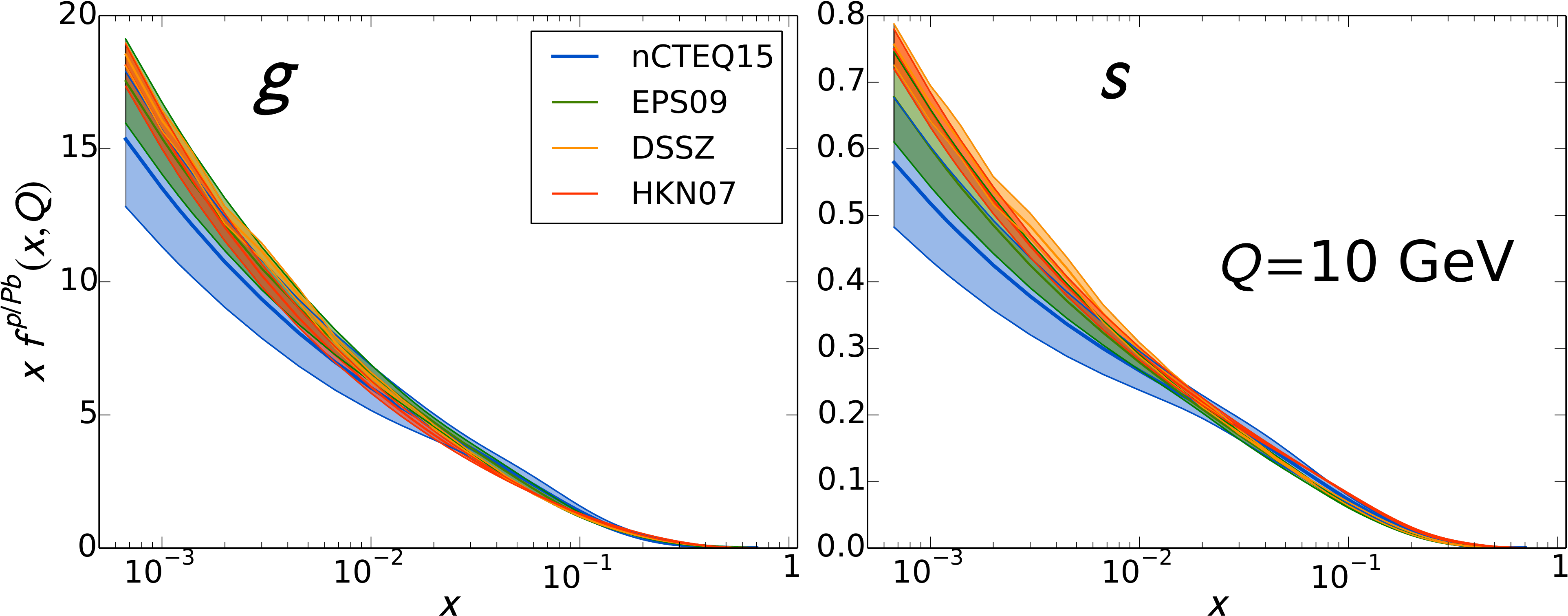}
\vspace{-0.05in}
\caption{Selected PDF flavors from  nCTEQ15~\cite{Kovarik:2015cma}  compared with 
 results from the literature~\cite{Hirai:2007sx,Eskola:2009uj,deFlorian:2011fp}.
}
\vspace{-0.10in}
\label{fig:gspdfs}
\end{figure} 

\begin{wrapfigure}{R}{0.45\textwidth} 
\centering{} 
\vspace{-30pt}
\includegraphics[width=0.45\textwidth]{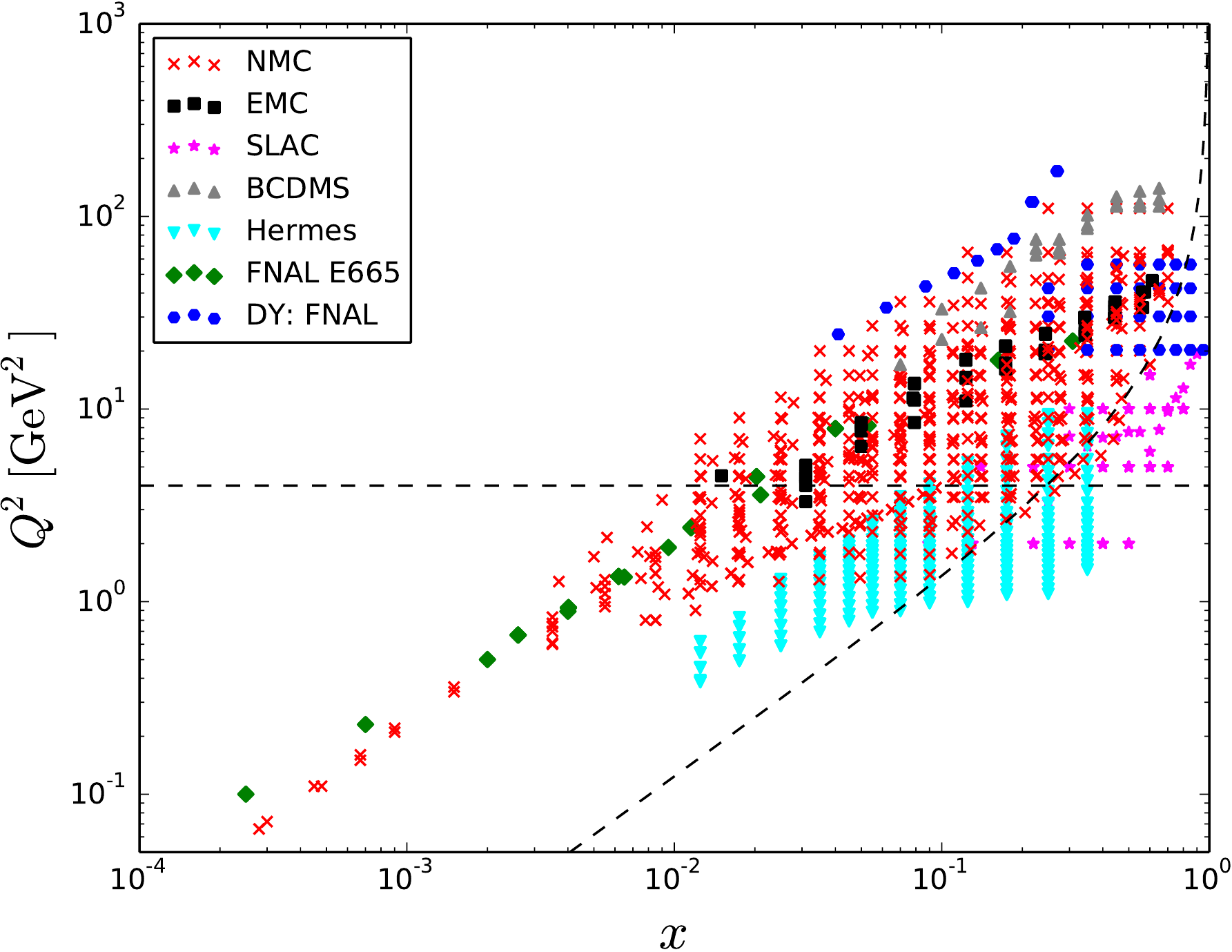}
\caption{
a)~Kinematic reach of DIS and DY data used in
the  nCTEQ15 fits~\cite{Kovarik:2015cma}. The dashed lines represent the
kinematic cuts:  ($Q > 2~{\rm GeV}$,
$W > 3.5~{\rm GeV}$). 
}
\label{fig:kin}
\end{wrapfigure}

\newpage
\section{The nCTEQ Project}

The  nCTEQ project\footnote{For details, see {\tt www.ncteq.org} which is hosted at HepForge.org.} 
built upon the work of the CTEQ proton PDF global fitting effort by 
extending the fit degrees of freedom into the nuclear dimension. 
Previous to the nCTEQ effort, nuclear data was ``corrected'' to isoscalar data 
and added to the proton PDF fit {\it without} any uncertainties. 
In contrast, the nCTEQ framework allows full communication between the nuclear data 
and the proton data, as illustrated schematically in Fig.~\ref{fig:scheme}.
For example, this enables us to investigate if observed tensions between
data sets could potentially be  attributed to the nuclear corrections. 

The details of the nCTEQ program is presented in Ref.~\cite{Kovarik:2015cma}. 
Figure~\ref{fig:gspdfs} displays selected flavors from the nCTEQ15 set,
and these  compare favorably to other determinations from the literature~\cite{Hirai:2007sx,Eskola:2009uj,deFlorian:2011fp}.
This analysis used  Deeply Inelastic Scattering (DIS), lepton pair production (Drell-Yan), and pion production 
from a variety of experiments totaling 740  data (after cuts) and 19 nuclei.
The kinematic range of the data in the $\{ x,Q^2\}$ plane is displayed in Fig.~\ref{fig:kin}.

\section{LHC Heavy Ion $W$ Production}

\begin{figure}[th] 
\centering{} 
\null \vspace{-0.2in}
\includegraphics[width=0.45\textwidth]{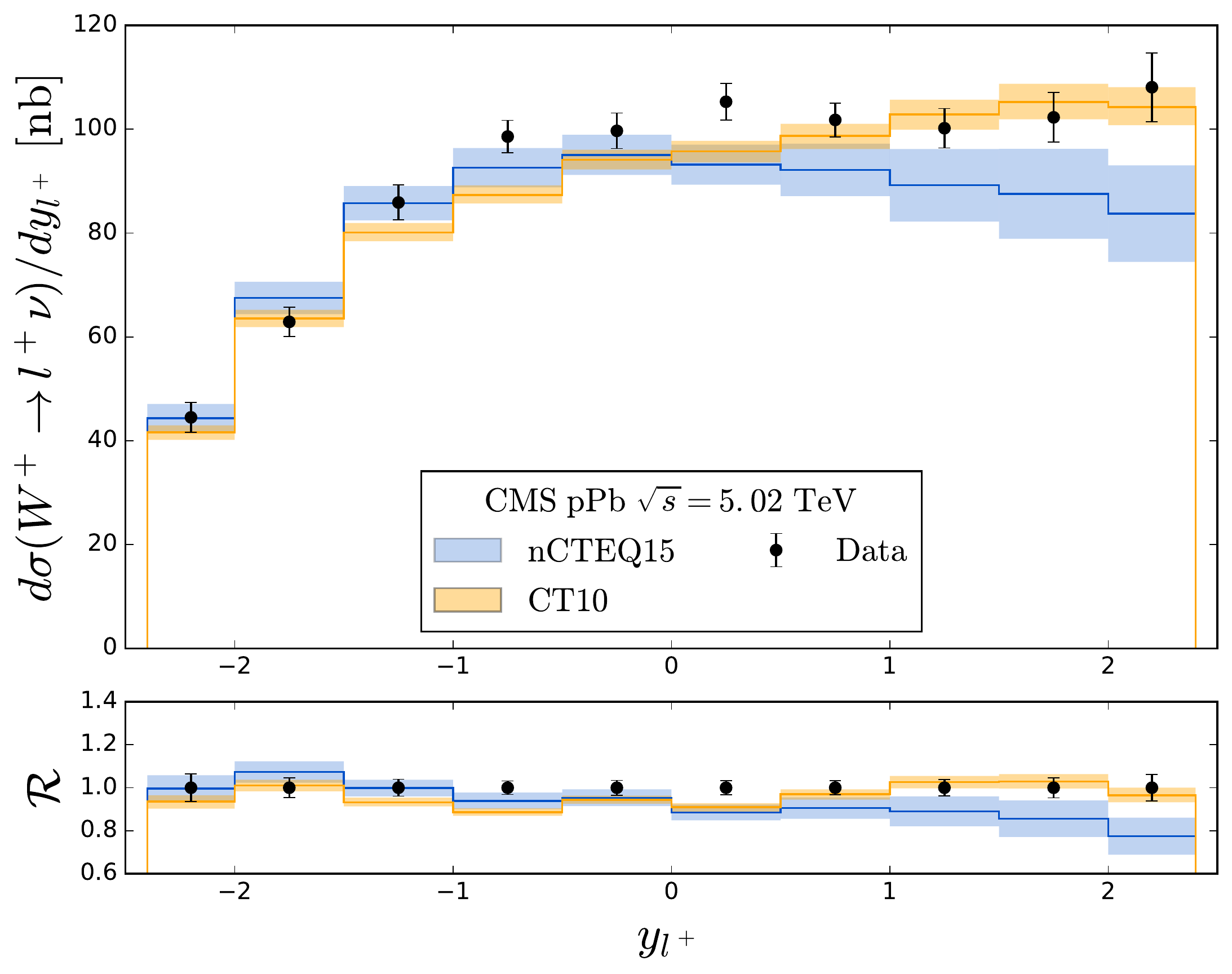}
\hfil
\includegraphics[width=0.45\textwidth]{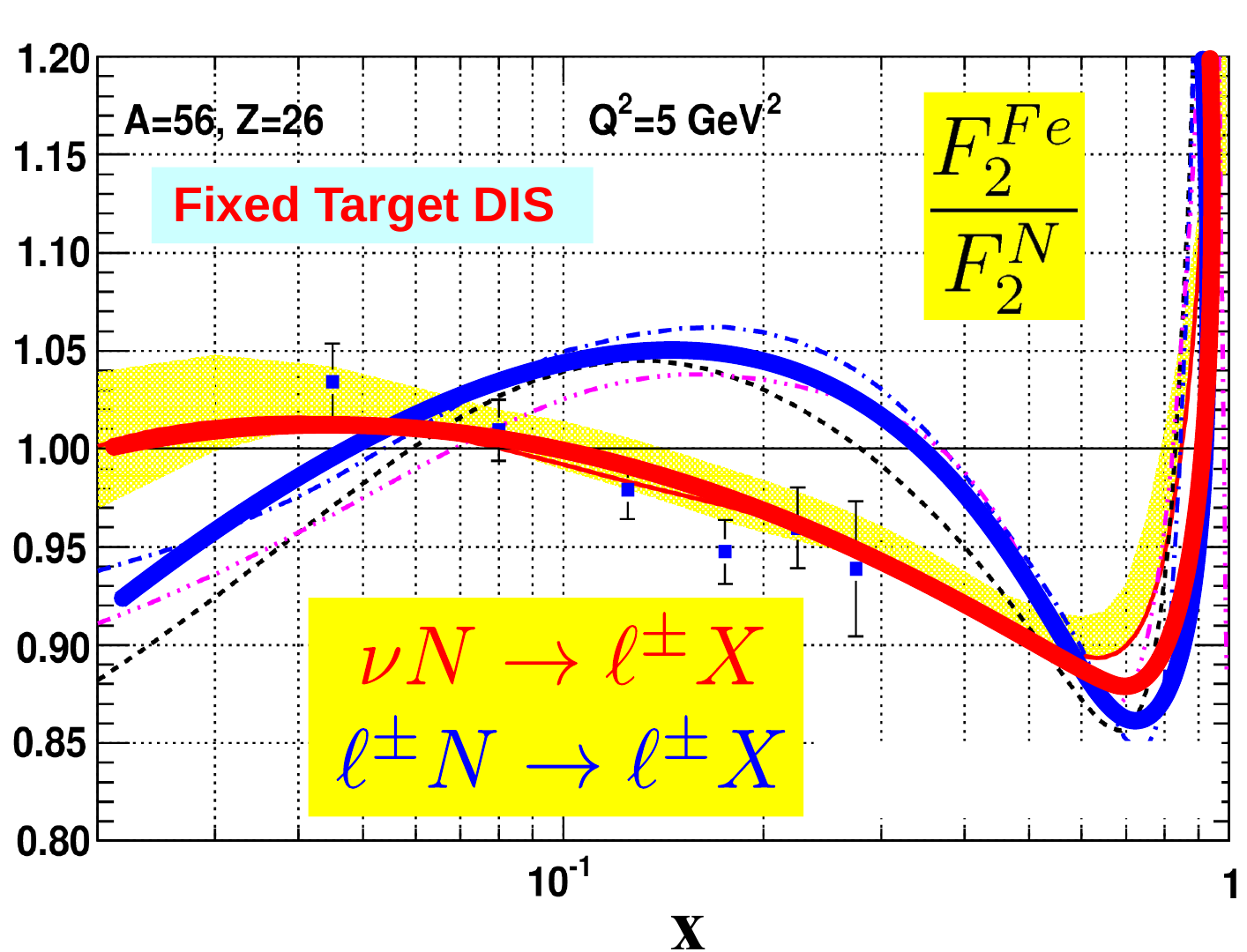}
\caption{
a)~CMS $W^+$  production in pPb collisions at the LHC~\cite{Kusina:2016fxy}.
\quad 
b)~The computed nuclear correction ratio,
$F_2^{Fe}/F_2^{D}$ as a function of $x$ for $Q^2=5~{\rm GeV}^2$, adapted from Ref.~\cite{Schienbein:2009kk}.
The  blue  line is for neutral current DIS charged-lepton scattering $\ell^\pm N \to \ell^\pm X$,
and 
the  red  line is for charged current DIS neutrino scattering $\nu N \to \ell^\pm X$.
}
\label{fig:rap}
\end{figure} 

Although the nCTEQ15 data covered an extensive range of kinematics and  nuclear A, 
there is new heavy ion data from the LHC using proton--lead (pPb) and lead--lead (PbPb) collisions 
at even higher energies. 
We shall be  interested to see how strongly this data can influence the nCTEQ15 PDFs. 
We will focus our discussion  on the LHC heavy ion  $W/Z$ rapidity  distributions\footnote{%
For a more extensive study, see Refs.~\cite{Kusina:2016fxy,Eskola:2016oht}.
}
as we expect this can provide information on the (less constrained) strange and charm PDFs. 
The justification  for this expectation  is presented  in  Ref.~\cite{Kusina:2012vh} 
which computes the relative partonic contributions for $W/Z$ production  
at the Tevatron ($p\bar{p}$) and the LHC ($pp$). 
As the $\sqrt{s}$ increases, the heavier quarks $\{s,c,b\}$ provide an increased 
contribution to $W/Z$ production.

In Fig.~\ref{fig:rap}-a) we display the CMS rapidity distribution for $W^+$ 
production in pPb collisions with $\sqrt{s}=5.02~{\rm TeV}$. 
We show the results using the nuclear PDFs from nCTEQ15 (with nuclear  corrections), and for comparison 
we also display  CT10 PDFs (without nuclear  corrections).

The kinematics of the process are defined  such that negative rapidity corresponds  to large lead $x$,
and positive rapidity is small lead $x$. 
Thus, as we scan across the distribution in rapidity, we effectively scan in $x$. 
This is particularly evident when we compare nCTEQ15  with the CT10 results. 
At  negative rapidity (large lead $x$) nCTEQ15 is above CT10, while 
at  positive rapidity (small lead $x$) nCTEQ15 is below.
We compare this pattern with the expected nuclear correction as shown in  Fig.~\ref{fig:rap}-b).
Focusing on the (blue) line for charged lepton DIS, we see the negative rapidity corresponds to the 
anti-shadowing region ($x \sim 0.2$) where the nuclear PDFs are larger than the proton PDFs ($F_2^{Fe}>F_2^{D}$),
while 
 the positive rapidity corresponds to the 
shadowing region ($x \lesssim 0.05$) where the nuclear PDFs are smaller than the proton PDFs ($F_2^{Fe}<F_2^{D}$).
This very direct mapping between rapidity and $x$ can help us to better constrain the PDF $x$ distribution. 

We now consider the comparison between the nCTEQ result  and the displayed data. 
At negative rapidity (large lead $x$) the nCTEQ15 result compares well with the data, 
but at  positive rapidity (small lead $x$) the nCTEQ15 result significantly undershoots the data. 
Examining Fig.~\ref{fig:kin}, we see that the bulk of the data used in nCTEQ15 (after cuts) is at relatively
large $x$ values. Thus, the good comparison between nCTEQ15 and the data at large $x$  is encouraging.
But the poor agreement at smaller $x$ suggests that the initial 740 data points used for the nCTEQ15  
may not have been sufficient  to constrain the small $x$ behavior; thus, including the new LHC heavy ion $W/Z$
data into the nPDF fit might help reduce the uncertainties in the small $x$ region.

There is an  additional interesting observation we can make regarding  Fig.~\ref{fig:rap}-a);
In  Fig.~\ref{fig:rap}-b), if we use the (red) neutrino DIS nuclear correction instead of the 
(blue) charged lepton result, the shadowing effect would be reduced  and the nCTEQ15 result would move toward
the data in the small $x$ region. 
The choice of which nuclear correction is correct has been  discussed extensively in the literature,
and there is no definitive resolution at present.\footnote{%
See Refs.~\cite{Kovarik:2010uv,Paukkunen:2013grz}, and also references therein.
}
Given that the LHC heavy ion $W/Z$ data is sensitive to the shadowing/anti-shadowing region, 
it could prove illuminating to include this into the fit.

\section{Cross Section Correlations}

\begin{wrapfigure}{R}{0.45\textwidth} 
\centering{} 
\vspace{-1.2in}
\includegraphics[width=0.45\textwidth]{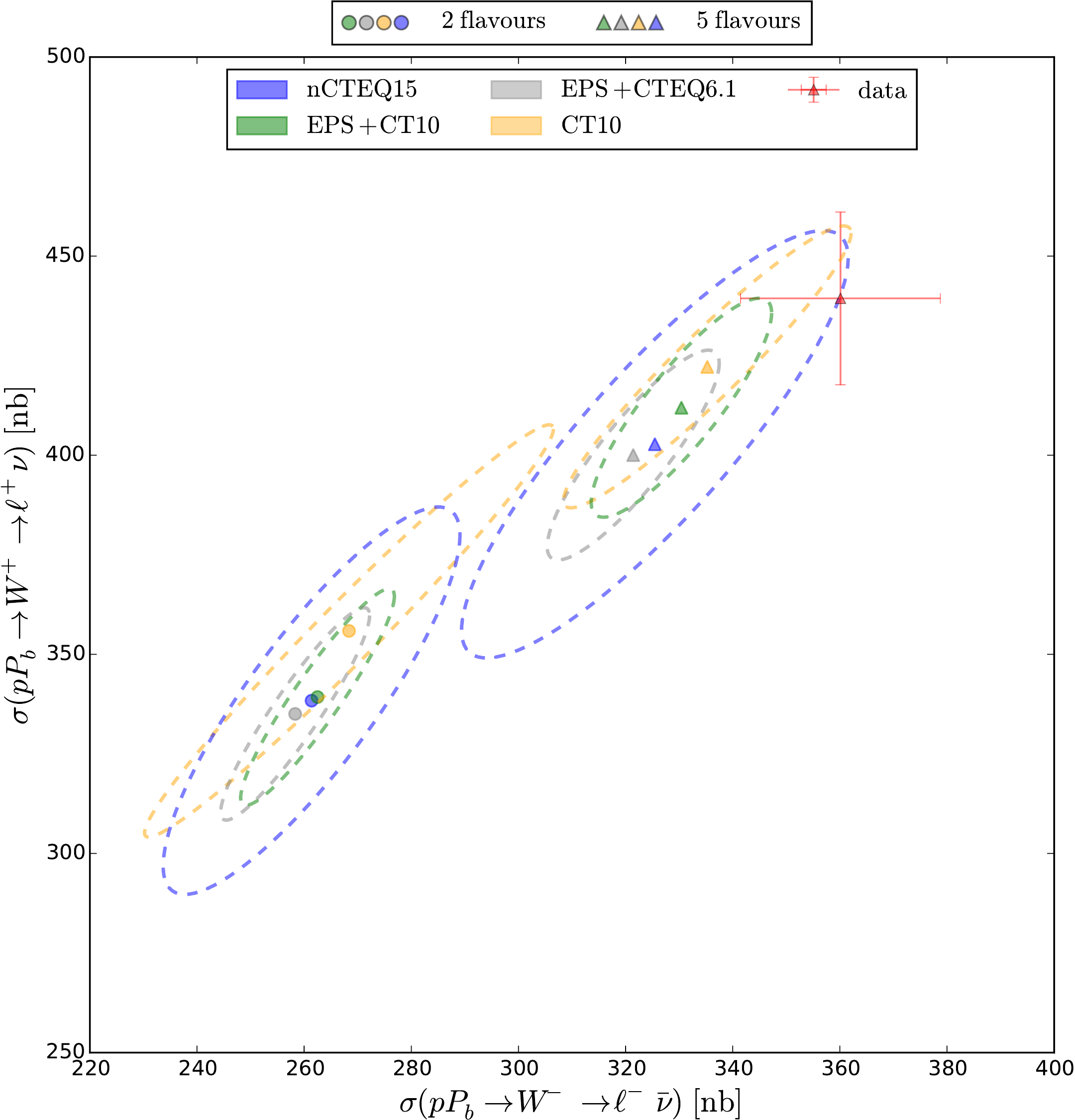}
\caption{Comparison of correlations between $W^+$ and
$W^-$ cross sections for the 
 5~flavor case ($\blacktriangle$) and
 2~flavor case ($\bullet$).
 We show  results for
nCTEQ15, EPS09+CT10, EPS09+CTEQ6.1 and CT10
PDFs overlaid with the CMS data~\cite{Kusina:2016fxy}.
}
\label{fig:cor}
\end{wrapfigure}

The previous analysis suggests that new LHC heavy ion data can influence the nCTEQ15 PDFs. 
We  would also like to specifically determine the impact of this data on the heavy flavors such as strange and charm. 
For this purpose, in Fig.~\ref{fig:cor} we display the correlation cross sections 
for $W^\pm$ for 
5 flavors of quarks $\{u,d,s,c,b\}$, and also for 
2 flavors of quarks $\{u,d\}$. 
The obvious shift of these cases indicates the contribution of 
the heavy flavors $\{s,c,b\}$ is substantial; hence, these data are sensitive  to, 
and can help constrain, the  parton flavors. 
For both the  5~flavor and 2~flavor cases, 
we display 4 results; 
there are 3 results with nuclear corrections\footnote{%
We have added  CTEQ6.1+EPS09 as CTEQ6.1 was the baseline used for the EPS09 fit.}
(nCTEQ15, CT10+EPS09,  CTEQ6.1+ EPS09),
and 1 result without nuclear corrections (CT10).

For the 5~flavor case, the calculations are scattered to the low side of the data 
in both $W^+$ and $W^-$.
The CT10 result is the closest to the data, but due to the larger uncertainties of nCTEQ15, 
the data point is  within range of both of their ellipses.
We also observe that the  CT10+EPS09 and CTEQ6.1+EPS09 results bracket the nCTEQ15 value;
this is due to the very different strange PDF associated with CT10 and CTEQ6.1.

For the 2~flavor case, all the nuclear results (nCTEQ15,
CT10+EPS09, CTEQ6.1+EPS09) coalesce, and they are distinct from the
non-nuclear result (CT10).
This pattern suggests that the nuclear corrections of nCTEQ15 and
EPS09 for the $\{u,d \}$ flavors are quite similar, and the spread
observed in the 5~flavor case comes from differences of $s(x)$ in the
underlying base PDF.\footnote{%
While the charm PDF does play a role in the above (the bottom
contribution is minimal), $c(x)$ is generated radiatively by the
process $g\to c\bar{c}$ (we assume no intrinsic component)~\cite{Lyonnet:2015dca}; thus, it
is essentially determined by the charm mass value and the gluon
PDF. In contrast, the “intrinsic” nature of the strange PDF leads to
its comparably large uncertainties. Therefore, the strange quark PDF
will be  primarily driving the observed  differences.}
Thus we infer that the difference between the nuclear results and the
proton result accurately represents the nuclear corrections for the
2~flavor case (for $\{u,d\}$), but for the 5~flavor case it is a mix
of nuclear corrections and variations of the underlying sea quarks.

\section{Conclusion}

In this brief report we have observed that the LHC heavy ion data can help
determine nuclear corrections for large A values in a 
kinematic $\{x,Q^2\}$ range very different from nuclear corrections
provided by fixed-target measurements.
The $W/Z$ \   pPb data are sensitive to the heavier quark flavors
(especially the strange PDF), so this provides important information
on the nuclear flavor decomposition.
Improved information on the nuclear corrections from the LHC lead data
can also help reduce proton PDF uncertainties as (at present) fixed-target nuclear
data is essential for distinguishing the individual flavors~\cite{Ball:2017nwa}.
The next step is to incorporate this new data into the nPDF fit to
help separately disentangle issues of flavor differentiation and nuclear corrections.


\begin{thebibliography}{10}

\bibitem{Kovarik:2015cma}
K.~Kovarik et~al. 
[The nCTEQ Collaboration],
\newblock {nCTEQ15 - Global analysis of nuclear parton distributions with
  uncertainties in the CTEQ framework}.
\newblock {\em Phys. Rev.}, D93(8):085037, 2016.

\bibitem{Aaboud:2016btc}
Morad Aaboud et~al.
\newblock {Precision measurement and interpretation of inclusive $W^+$ , $W^-$
  and $Z/\gamma ^*$ production cross sections with the ATLAS detector}.
\newblock {\em Eur. Phys. J.}, C77(6):367, 2017.

\bibitem{Hirai:2007sx}
M.~Hirai, S.~Kumano, and T.~H. Nagai.
\newblock {Determination of nuclear parton distribution functions and their
  uncertainties in next-to-leading order}.
\newblock {\em Phys. Rev.}, C76:065207, 2007.

\bibitem{Eskola:2009uj}
K.~J. Eskola, H.~Paukkunen, and C.~A. Salgado.
\newblock {EPS09: A New Generation of NLO and LO Nuclear Parton Distribution
  Functions}.
\newblock {\em JHEP}, 04:065, 2009.

\bibitem{deFlorian:2011fp}
Daniel de~Florian, Rodolfo Sassot, Pia Zurita, and Marco Stratmann.
\newblock {Global Analysis of Nuclear Parton Distributions}.
\newblock {\em Phys. Rev.}, D85:074028, 2012.

\bibitem{Schienbein:2009kk}
I.~Schienbein, J.~Y. Yu, K.~Kovarik, C.~Keppel, J.~G. Morfin, F.~Olness, and
  J.~F. Owens.
\newblock {PDF Nuclear Corrections for Charged and Neutral Current Processes}.
\newblock {\em Phys. Rev.}, D80:094004, 2009.

\bibitem{Kusina:2016fxy}
A.~Kusina, et al. 
[The nCTEQ Collaboration],
\newblock {Vector boson production in proton-lead and lead-lead collisions at
  the LHC and its impact on nCTEQ15 PDFs}.
\newblock {\em Eur. Phys. J.}, C77(7):488, 2017.


\bibitem{Eskola:2016oht}
Kari~J. Eskola, Petja Paakkinen, Hannu Paukkunen, and Carlos~A. Salgado.
\newblock {EPPS16: Nuclear parton distributions with LHC data}.
\newblock {\em Eur. Phys. J.}, C77(3):163, 2017.

\bibitem{Kusina:2012vh}
A.~Kusina, T.~Stavreva, S.~Berge, F.~I. Olness, I.~Schienbein, K.~Kovarik,
  T.~Jezo, J.~Y. Yu, and K.~Park.
\newblock {Strange Quark PDFs and Implications for Drell-Yan Boson Production
  at the LHC}.
\newblock {\em Phys. Rev.}, D85:094028, 2012.

\bibitem{Kovarik:2010uv}
K.~Kovarik, I.~Schienbein, F.~I. Olness, J.~Y. Yu, C.~Keppel, J.~G. Morfin,
  J.~F. Owens, and T.~Stavreva.
\newblock {Nuclear corrections in neutrino-nucleus DIS and their compatibility
  with global NPDF analyses}.
\newblock {\em Phys. Rev. Lett.}, 106:122301, 2011.

\bibitem{Paukkunen:2013grz}
Hannu Paukkunen and Carlos~A. Salgado.
\newblock {Agreement of Neutrino Deep Inelastic Scattering Data with Global
  Fits of Parton Distributions}.
\newblock {\em Phys. Rev. Lett.}, 110(21):212301, 2013.

\bibitem{Lyonnet:2015dca}
Florian Lyonnet, Aleksander Kusina, Tomáš Ježo, Karol Kovarík, Fred Olness,
  Ingo Schienbein, and Ji-Young Yu.
\newblock {On the intrinsic bottom content of the nucleon and its impact on
  heavy new physics at the LHC}.
\newblock {\em JHEP}, 07:141, 2015.


\bibitem{Ball:2017nwa} 
  R.~D.~Ball {\it et al.} [NNPDF Collaboration],
\newblock {Parton distributions from high-precision collider data},
\newblock {\em Eur. Phys. J.}, C77(10):663, 2017.


\end{thebibliography}
%



\end{document}